\documentclass[letter,
               keeplastbox,   
               ]{jacow}
\usepackage{pdfpages,multirow,ragged2e} %
\makeatletter%
	\ifboolexpr{bool{xetex}}
	 {\renewcommand{\Gin@extensions}{.pdf,%
	                    .png,.jpg,.bmp,.pict,.tif,.psd,.mac,.sga,.tga,.gif,%
	                    .eps,.ps,%
	                    }}{}
\makeatother

\ifboolexpr{bool{xetex} or bool{luatex}} 
 {}                                      
 {\usepackage[utf8]{inputenc}}           

\usepackage[USenglish]{babel}

\usepackage{amssymb}
\usepackage{amsmath}
\usepackage{geometry} 
\begin{document}

\title{Circular Modes for Linacs\thanks{\NoCaseChange{This work was supported by the U.S. Department of Energy, under Contract No. DE-AC02-06CH11357.}}}
 
\author{Onur Gilanliogullari\thanks{ogilanli@hawk.iit.edu}, Pavel Snopok, Illinois Institute of Technology, Chicago, IL, USA \\
		Brahim Mustapha, Argonne National Laboratory, Lemont, IL, USA \\
		}	
\maketitle

\begin{abstract}
    Circular mode beams are beams with non-zero angular momentum and strong inter-plane coupling. This coupling can be achieved in linear accelerators (linacs) through magnetization of electrons or ions at the source. Depending on the magnetization strength, the intrinsic eigenmode emittance ratio can be large, which produces intrinsic flatness. This flatness can either be converted to real space flatness or can be maintained as round coupled beam through the system. In this paper, we discuss rotation invariant designs that allow circular modes to be transported through the lattice while accelerating the beam and maintaining its circularity. We demonstrate that with rotation invariant designs the circularity of the mode can be preserved as round beam while maintaining intrinsic flatness to be converted to flat beam for high brightness or injected into a ring.
\end{abstract}

\section{Introduction}
Space charge effects at low energies are the limiting factor for achieving high-intensity high-quality beams. This limit originates from two sources. Firstly, the current limit due to extraction at the source. Secondly, the space charge limit due to self Coulomb repulsion force and charge density within the beam. 
Space charge forces are equivalent in the transverse plane when the beam is round. When the beam is flat, the quality parameters such as beam brightness and luminosity are enhanced due to small geometric beam size in one dimension. At the same time, flat beams can not maintain high currents due to space charge effect, which tends to equilibrate the beam sizes. Circular modes are round beams with intrinsic flatness and non-zero angular momentum as discussed in~\cite{gilanliogullaricircmode,burov2002circular}. 

Since space charge forces depend on the beam distribution in space, circular modes behave in a similar fashion to uncorrelated round beams while maintaining their intrinsic flatness. 

Magnetized beams can also be classified as circular modes, where the beam experiences only one part of the solenoid's fringe field, which translates the canonical angular momentum to mechanical angular momentum. Extension of Bush's theorem discusses the angular momentum conversion from the magnetic field strength as discussed in ~\cite{groening2018extension}.

In this paper, we will provide the background and define circular modes through coupling vectors. We will discuss formation of circular modes at the source and present rotational-invariant linac lattice designs.

\section{Circular Mode Optics}

Due to coupling, phase space vector parameterization is expressed through coupled optics as shown in~\cite{lebedev2010betatron}. Phase space vector, $\Vec{z}=[x,x',y,y']^{T}$, can be written as
\begin{equation}
    \Vec{z} = \frac{1}{2}\sqrt{\epsilon_{1}}\Vec{v}_{1}e^{i\psi_{1}} + \frac{1}{2}\sqrt{\epsilon_{2}}\Vec{v}_{2}e^{i\psi_{2}} + c.c.,
\end{equation}
where $\epsilon_{1,2}$ are the eigenmode emittances, $\psi_{1,2}$ the betatron phases, $c.c.$ stands for complex conjugate, and $\vec{v}_{1,2}$ are the eigenvectors given by
\begin{equation}
    \vec{v}_{1} = \begin{pmatrix}
        \sqrt{\beta_{1x}} \\
        -\frac{i(1-u) + \alpha_{1x}}{\sqrt{\beta_{1x}}} \\
        \sqrt{\beta_{1y}}e^{i\nu_{1}} \\
        -\frac{iu + \alpha_{1y}}{\sqrt{\beta_{1y}}}
    \end{pmatrix}, \qquad \vec{v}_{2}= \begin{pmatrix}
        \sqrt{\beta_{2x}}e^{i\psi_{2}} \\
        -\frac{iu + \alpha_{2x}}{\sqrt{\beta_{2x}}} \\
        \sqrt{\beta_{2y}} \\
        -\frac{(1-u)i + \alpha_{2y}}{\sqrt{\beta_{2y}}}
    \end{pmatrix}.
\end{equation}
Here $\beta_{1x},\beta_{1y},\beta_{2x},\beta_{2y}$ are the coupled betatron functions, $\alpha_{1x},\alpha_{1y},\alpha_{2x},\alpha_{2y}$ the alpha functions, $u$ the strength of coupling and $\nu_{1,2}$ the phases of coupling. As mentioned earlier, circular mode beams are intrinsically flat, $\epsilon_{2}\ll\epsilon_{1}$ with flatness ratio defined as $\mathcal{R}=\epsilon_{1}/\epsilon_{2}$. Due to intrinsic flatness, mode~2 contribution to phase space is small and the hase space vector can be written as
\begin{equation}
    \begin{split}
        x&= \sqrt{\epsilon_{1}\beta_{1x}}\cos\psi_{1}, \quad y=\sqrt{\epsilon_{1}\beta_{1y}}\cos(\psi_{1}-\nu_{1}), \\
        x'&= \sqrt{\frac{\epsilon_{1}}{\beta_{1x}}}(-\alpha_{1x}\cos\psi_{1} - (1-u)\sin\psi_{1}) \\
        y'&= \sqrt{\frac{\epsilon_{1}}{\beta_{1y}}}(-\alpha_{1y}\cos(\psi_{1}-\nu_{1}) - u\sin(\psi_{1}-\nu_{1})).
    \end{split}
    \label{eq:mode1cont}
\end{equation}
The phase of coupling determines the shapes of beam projections onto the different phase space planes. The projection onto real space is given by
\begin{equation}
    \begin{split}
        & \frac{x^{2}}{2\sigma_{x}^{2}} - \frac{2\Tilde{\alpha}}{\sigma_{x}\sigma_{y}}xy + \frac{y^{2}}{2\sigma_{y}^{2}} = 1 - \Tilde{\alpha}^{2}, \\
        & \Tilde{\alpha} = \frac{\epsilon_{1}\sqrt{\beta_{1x}\beta_{1y}}\cos\nu_{1} + \epsilon_{2}\sqrt{\beta_{2x}\beta_{2y}}\cos\nu_{2}}{\sqrt{\epsilon_{1}\beta_{1x}+\epsilon_{2}\beta_{2x}}\sqrt{\epsilon_{1}\beta_{1y} + \epsilon_{2}\beta_{2y}}},
    \end{split}
    \label{eq:projectionxygen}
\end{equation}
where the beam distribution can be tilted in the $(x,y)$ space. Disregarding mode~2 contribution will simplify the projection equation to
\begin{equation}
    \frac{x^{2}}{2\sigma_{x}^{2}} - \frac{2xy}{\sigma_{x}\sigma_{y}}\cos\nu_{1} + \frac{y^{2}}{2\sigma_{y}^{2}} = \sin^{2}\nu_{1}.
    \label{eq:projxycircmode}
\end{equation}
In order to satisfy round beam condition, the phase of coupling needs to be $\pi/2$ and the optics functions, $\beta_{1x}=\beta_{1y}=\beta_{0}$. Furthermore, based on circular mode properties, the strength of coupling is determined to be $u=1/2$. This leads to interesting relations, such as
\begin{itemize}
    \item Coupled beta functions of the same type, $\beta_{1x},\beta_{2x}$ and $\beta_{1y},\beta_{2y}$ are half of the uncoupled beta functions.
    \item Apparent RMS emittances are half of the eigenmode~1 emittance, $\epsilon_{1}=2\epsilon_{x,y}$.
\end{itemize}
Based on these properties, the intrinsic flatness of a circular mode beam is mapped to the $(x,y')$ and $(y,x')$ phase space planes, which gives rise to non-zero angular momentum.

Since the phase of coupling is responsible for determining the beam projections onto the different phase space planes, controlling the phase of coupling is important. The propagation of mode~1 and 2 phases of coupling is given by
\begin{equation}
    \begin{split}
        \frac{d\nu_{1}}{ds}= \frac{1-u}{\beta_{1x}} - \frac{u}{\beta_{1y}} - \frac{R}{2}\left(\sqrt{\frac{\beta_{1y}}{\beta_{1x}}} - \sqrt{\frac{\beta_{1x}}{\beta_{1y}}}\right)\sin\nu_{1}, \\
        \quad \frac{d\nu_{2}}{ds} = \frac{1-u}{\beta_{2y}} - \frac{u}{\beta_{2x}} - \frac{R}{2}\left(\sqrt{\frac{\beta_{2x}}{\beta_{2y}}} - \sqrt{\frac{\beta_{2y}}{\beta_{2x}}}\right)\sin\nu_{2}.
    \end{split}
    \label{Eq:phaseofcouplingprop}
\end{equation}
For coupling strength, $u=1/2$, and equal beta functions, the equations simplify to
\begin{equation}
    \begin{split}
        \frac{d\nu_{1}}{ds} &= \frac{1}{2\beta_{1x}} - \frac{1}{2\beta_{1y}}, \quad \frac{d\nu_{2}}{ds} = \frac{1}{2\beta_{2y}} - \frac{1}{2\beta_{2x}}.
    \end{split}
    \label{eq:simplifiedpropnu}
\end{equation}
Therefore, an important criteria in designing rotation-invariant optics is the proper propagation and preservation of the phase of coupling.

\section{Creation of Circular Modes}
\begin{figure}[tbp]
    \centering
    \includegraphics[width=0.49\linewidth]{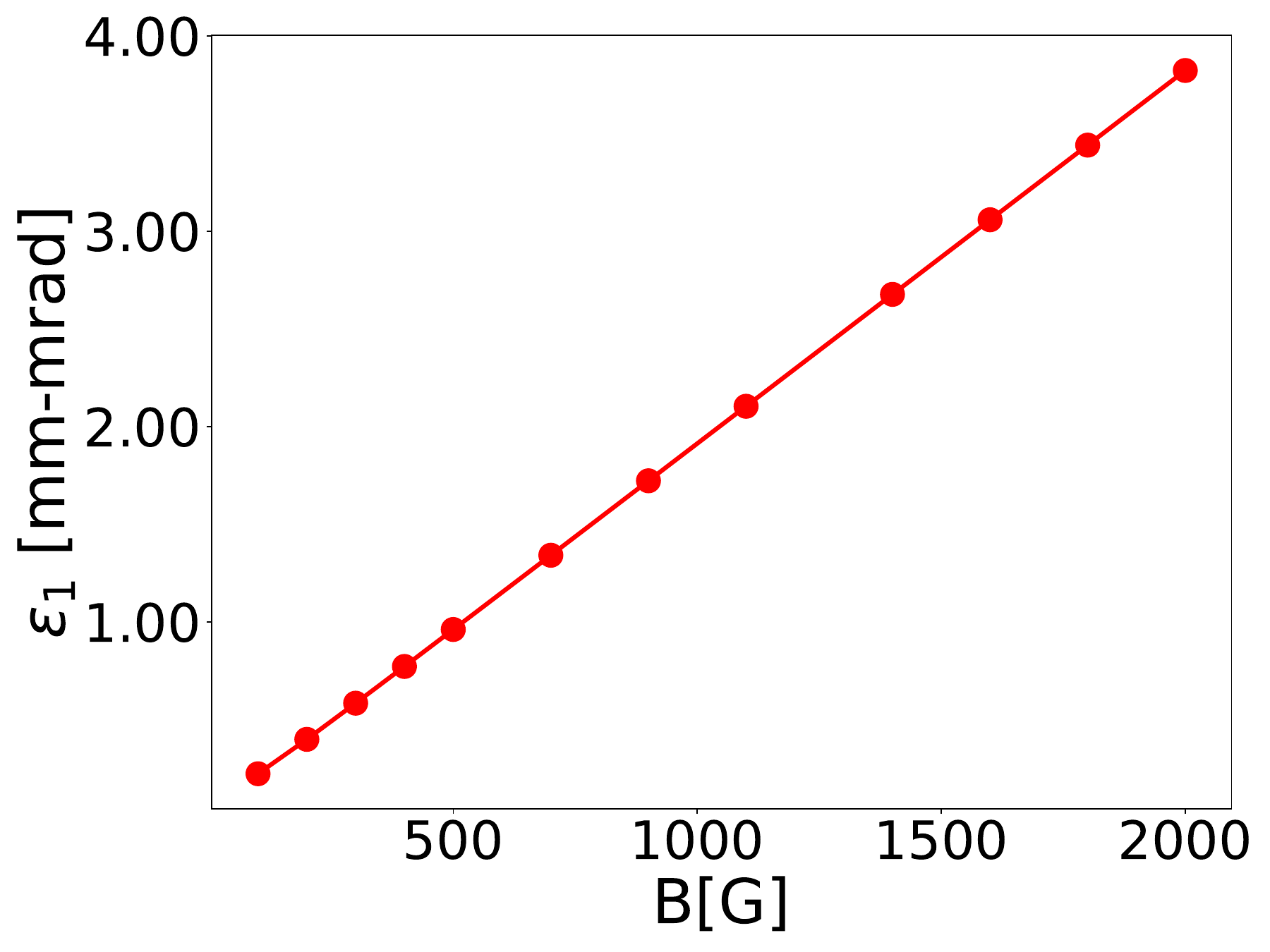}
    \includegraphics[width=0.49\linewidth]{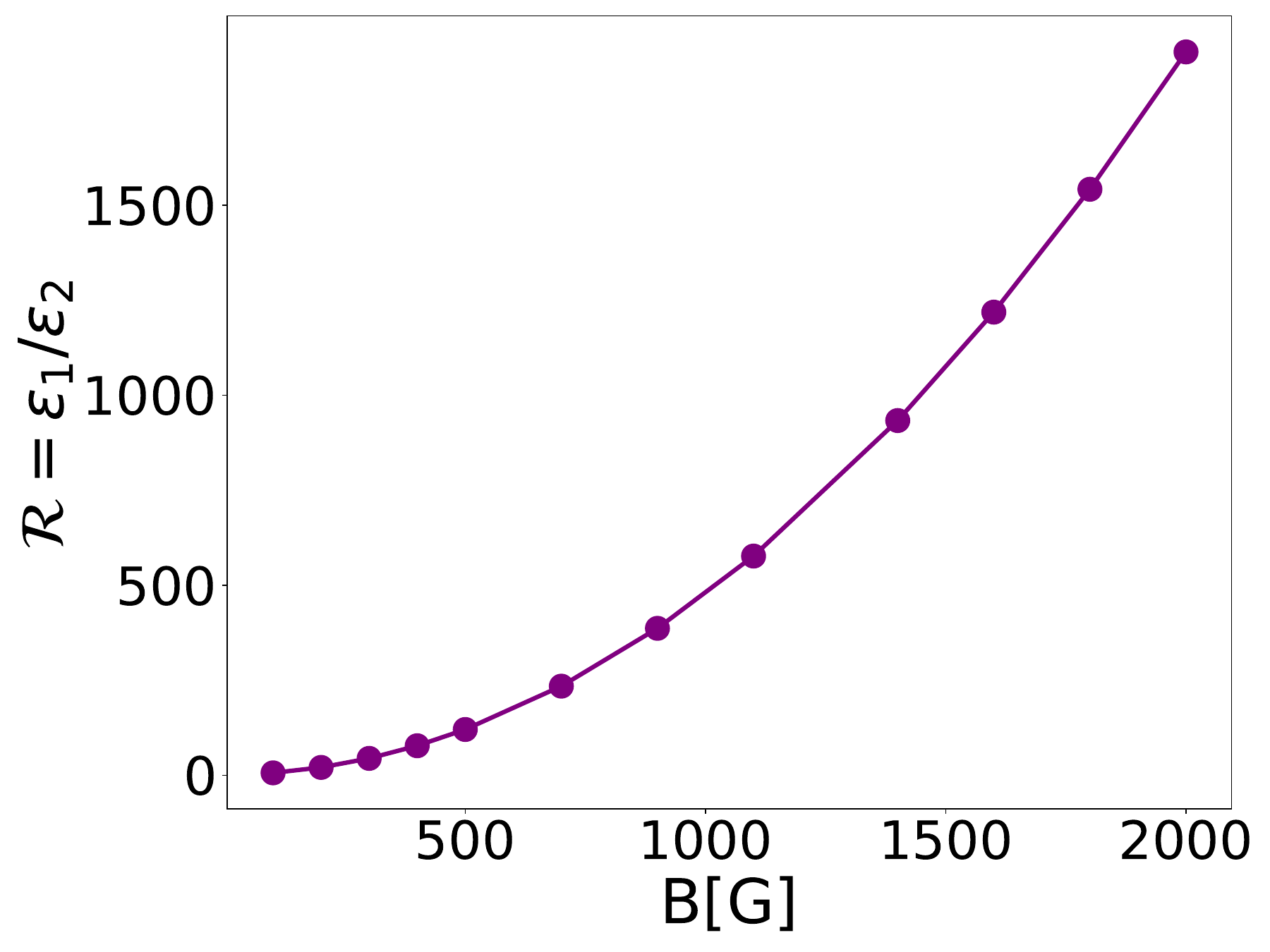}
    \includegraphics[width=0.49\linewidth]{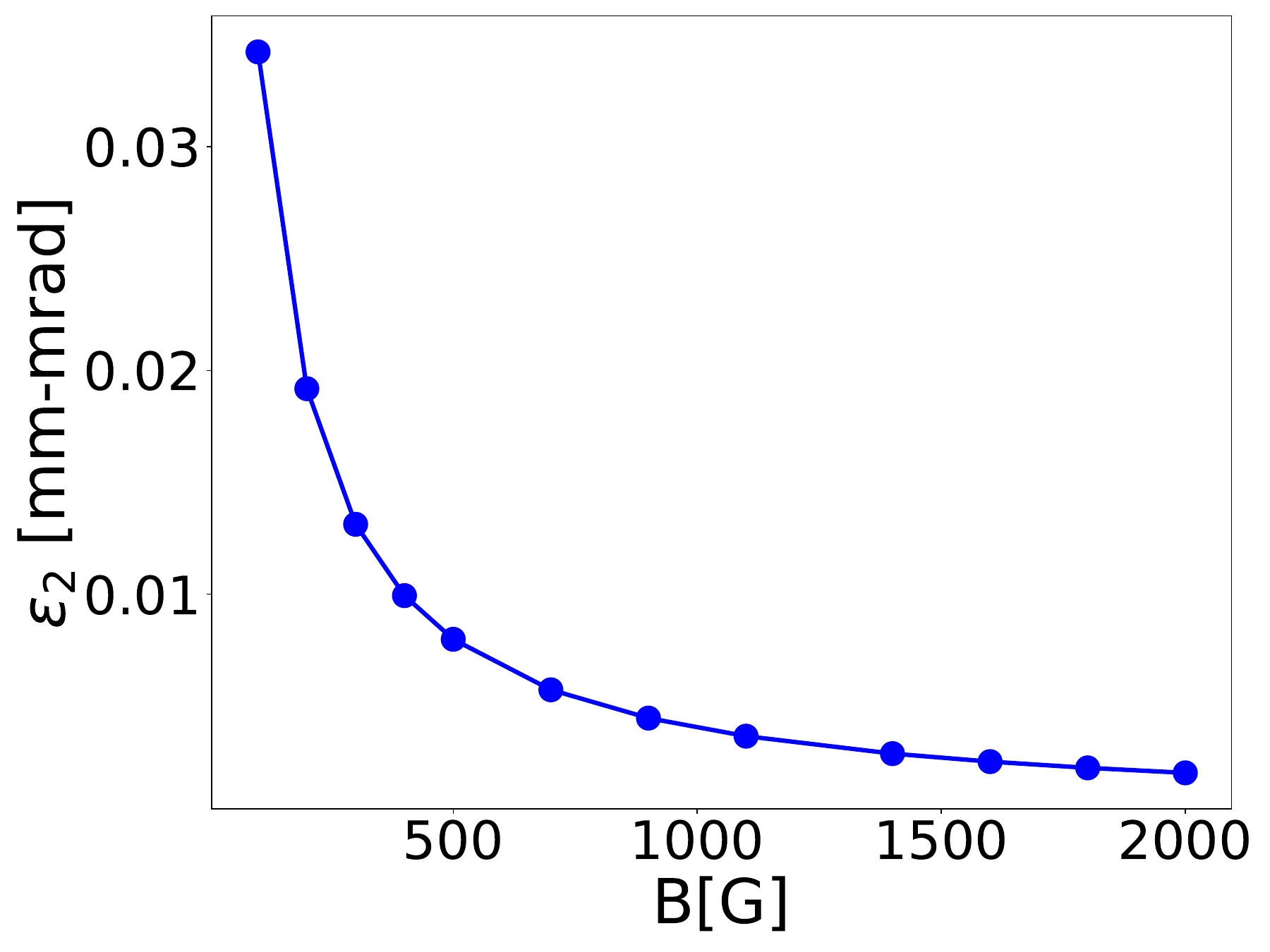}
    \includegraphics[width=0.49\linewidth]{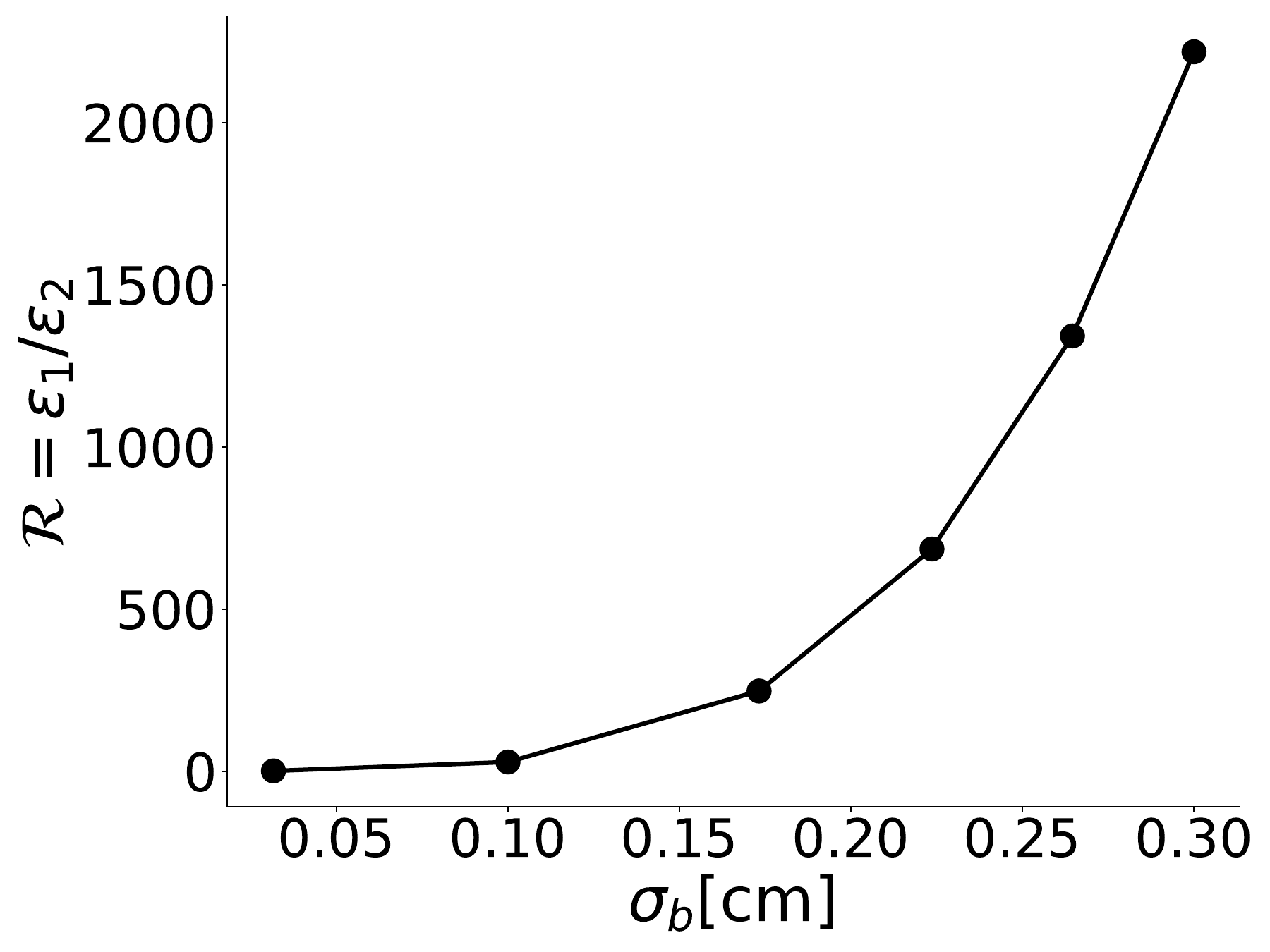}
    \caption{Magnetization of proton beams: eigenmode emittance~1 vs magnetic field strength (top-left), ratio of flatness vs magnetic field (top-right), eigenmode emittance~2 vs magnetic field (bottom-left), ratio of flatness vs beam size (bottom-right).}
    \label{fig:magnetizationgraphs}
\end{figure}

Magnetized beams, born inside a longitudinal magnetic field (solenoid), can be classified as circular mode beams. In this case, the beam particles experience only one side of the fringe field, given by the following matrix in thin lens approximation:
\begin{equation}
    \mathcal{M}_{\textnormal{fringe}} = \begin{pmatrix}
        1 & 0 & 0 & 0  \\
        0 & 1 & k & 0 \\
        0 & 0 & 1 & 0 \\
        -k & 0 & 0 & 1
    \end{pmatrix}.
\end{equation}
Here, $\displaystyle{k=\frac{B_{s}}{2B\rho}}$ is the solenoid focusing strength. Transformation of the beam through the fringe field with small divergence yields a vortex condition:
\begin{equation}
    \begin{pmatrix}
        y \\
        y'
    \end{pmatrix}_{f} = \begin{pmatrix}
        0 & 1/k \\
        -k & 0
    \end{pmatrix}\cdot \begin{pmatrix}
        x \\
        x'
    \end{pmatrix}_{f},
\end{equation}
where subscript $f$ refers to coordinates after transformation. One can use Bush's theorem to calculate the angular momentum value based on the magnetic field strength. Knowing the initial 4D emittance, one can calculate the eigenmode emittances as
\begin{equation}
    \begin{split}
        \epsilon_{1}&=\frac{1}{2}L_{z} + \frac{1}{2}\sqrt{L_{z}^{2}+4\epsilon_{4D}}, \\
        \epsilon_{2}&=-\frac{1}{2}L_{z} + \frac{1}{2}\sqrt{L_{z}^{2}+4\epsilon_{4D}}.
    \end{split}
\end{equation}
Based on this, the magnetization process with varied magnetic field and initial beam size is given in Fig.~\ref{fig:magnetizationgraphs}.
As eigenmode~1 grows with the magnetic field, eigenmode~2 emittance asymptotically reaches zero, hence increasing the ratio of flatness. The plot of flatness ratio versus beam size for a fixed magnetic field suggests that, instead of increasing the magnetic field, one can increase the initial 4D emittance  to achieve a similar result . Enhancement in flatness ratio is a direct consequence of the fringe field effect, which preserves the beam size after transformation, $x_{i}=x_{f}$ and $y_{i}=y_{f}$, while $x'_{f}= ky_{i}$ and $y'_{f}=-kx_{i}$ indicate the rotation speed is increased through either magnetic field strength, $k$, or initial beam sizes, $\sigma_x$ and $\sigma_y$.

\section{Rotational Invariance}
Rotation invariant designs aim to minimize the oscillation of the phase of coupling as shown in Eq.~\eqref{Eq:phaseofcouplingprop}. Solenoids are a solid choice in conserving angular momentum and the phase of coupling as long as the coupled beta functions are equal at solenoid's location. RF cavities also conserve angular momentum with their transverse magnetic field being minimal. Acceleration causes a damping effect, which can be seen from Hill's equation:
\begin{equation}
	x'' + \frac{(\gamma\beta)'}{\gamma\beta}x' + \kappa_{x}x=0 , \quad
	y'' + \frac{(\gamma\beta)'}{\gamma\beta}y' + \kappa_{y}y=0.
	\label{Eq:Hillsequationwithacc}
\end{equation}
Here $\gamma,\beta$ are Lorentz relativistic factors, $\kappa_x$, $\kappa_y$ the periodic focusing functions. Using Hill's equation, one can show that for rotational invariant designs angular momentum is conserved, $\displaystyle{\frac{dL_{z}}{ds}=0}$. Redefining the coordinates as $\displaystyle{\tilde{x},\tilde{y}=\frac{x,y}{\sqrt{\beta\gamma}}}$ leads to redefining the angular momentum as normalized angular momentum, $L_{z,N}=\gamma\beta L_{z}$. As a result of this, normalized angular momentum is invariant under acceleration, which is a similar effect as normalized emittance.

\section{Results}
\begin{figure}[tbp]
    \centering
    \includegraphics[width=0.49\linewidth]{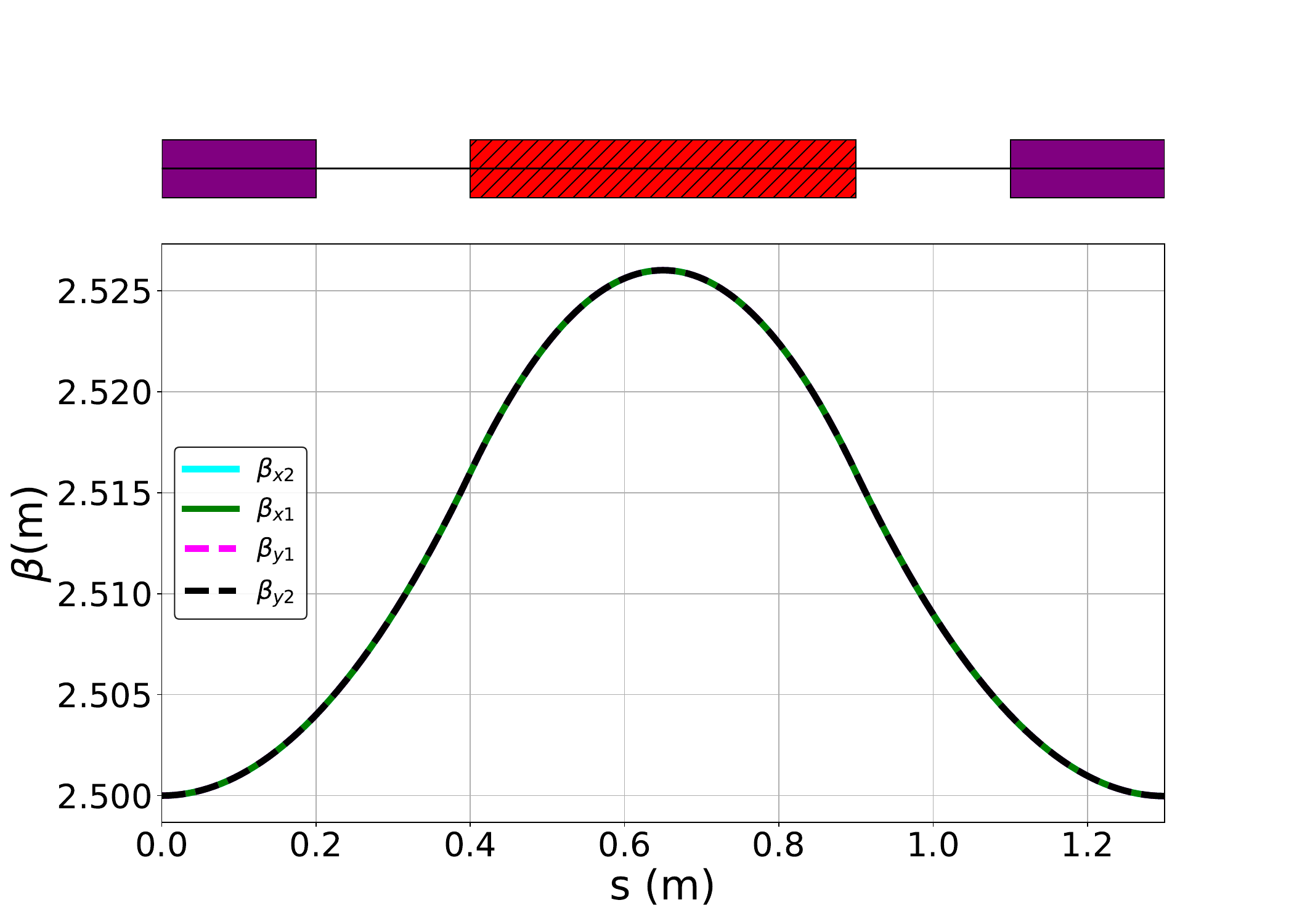}
    \includegraphics[width=0.49\linewidth]{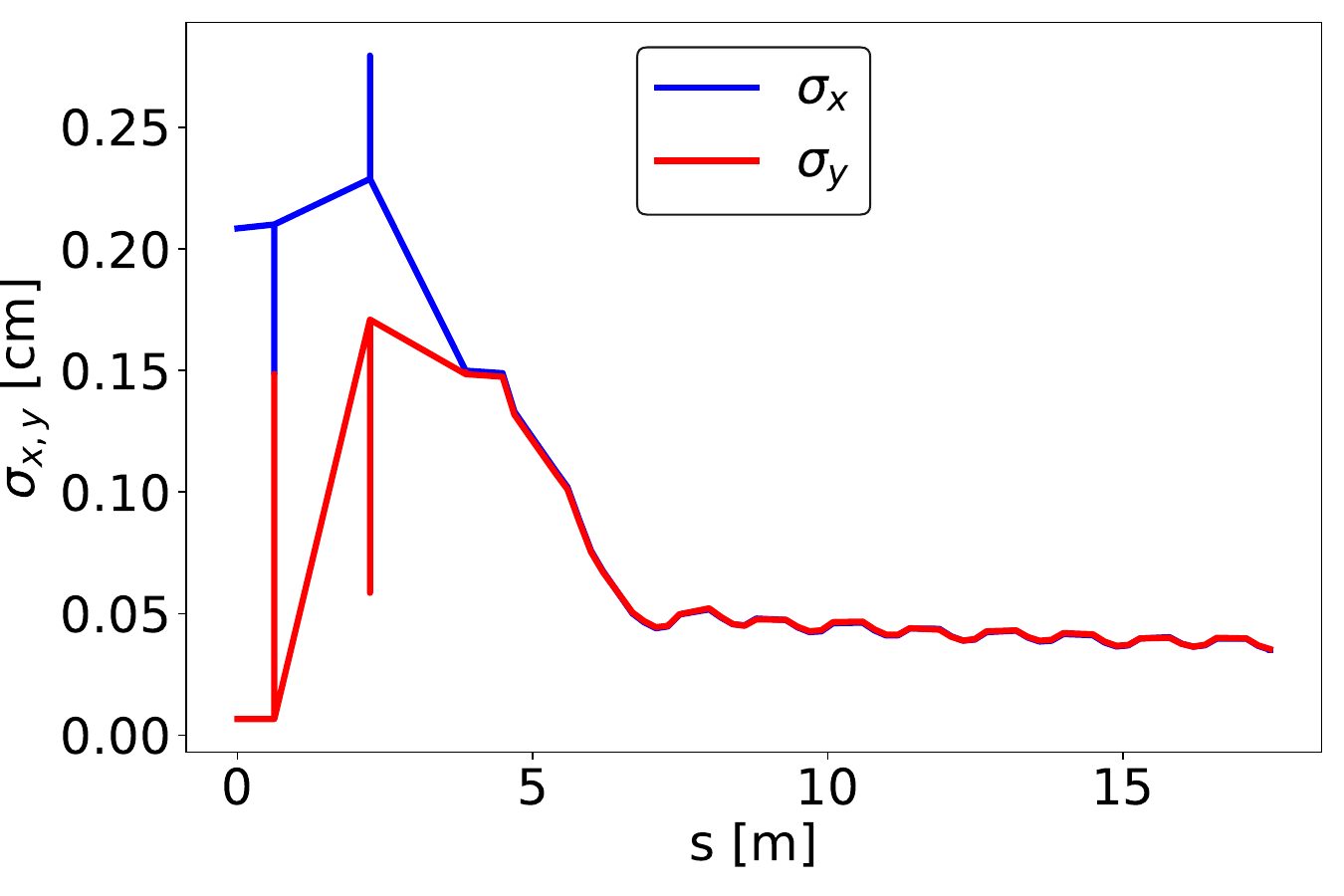}
    \includegraphics[width=0.49\linewidth]{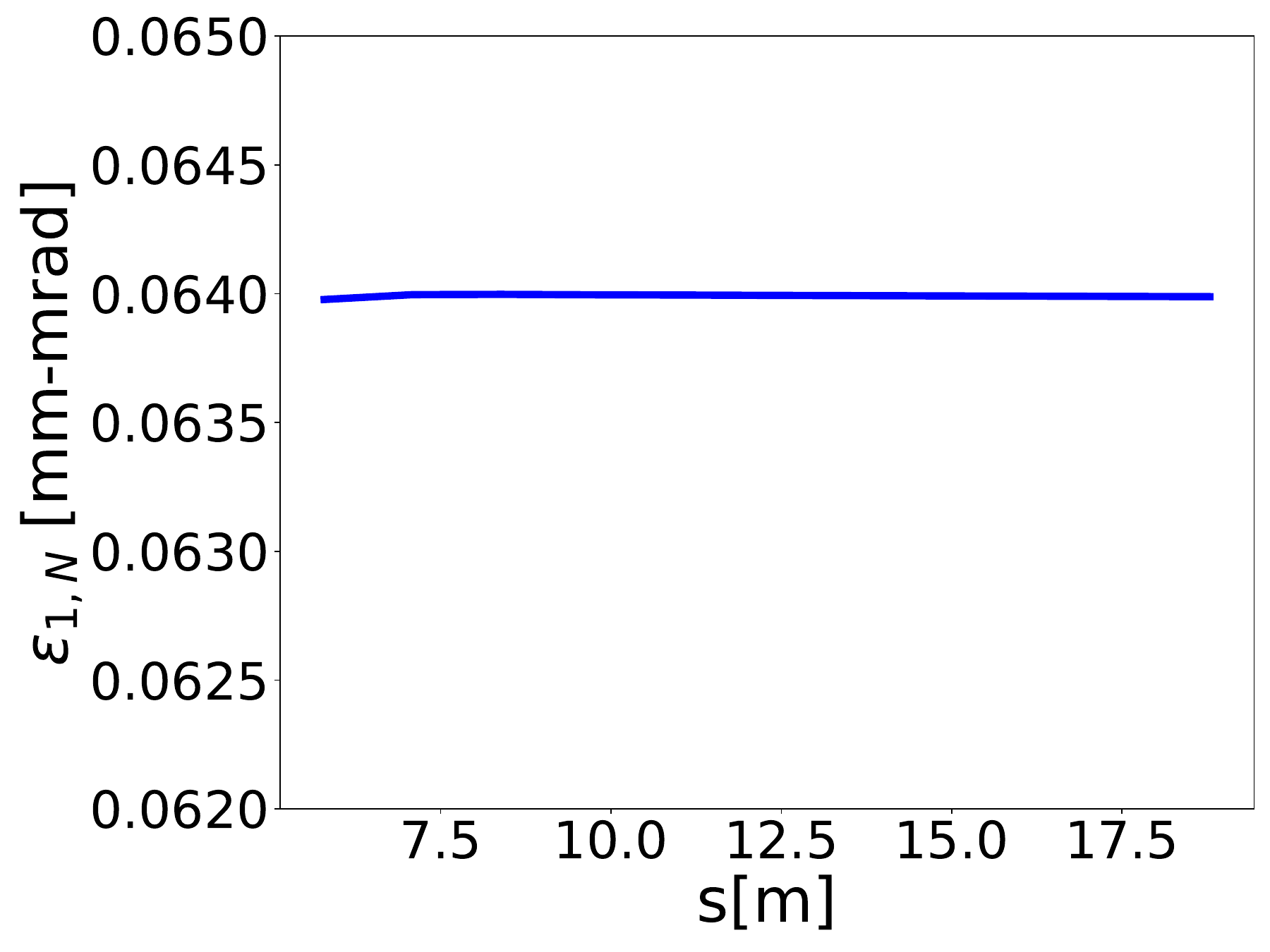}
    \includegraphics[width=0.49\linewidth]{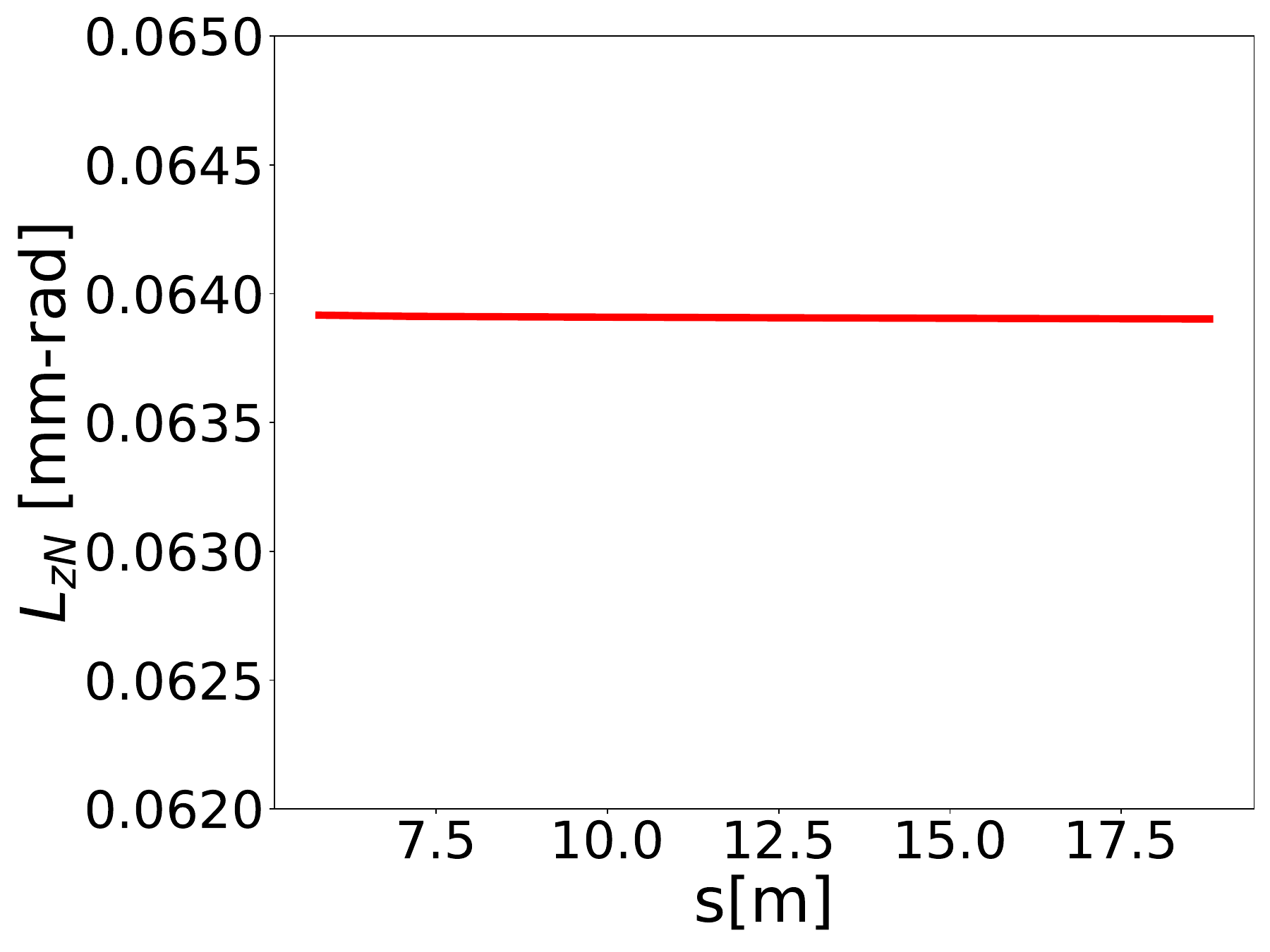}
    \caption{Repeating cell containing a solenoid between RF cavities (top-left), RMS envelopes (top-right), normalized eigenmode emittance~1 (bottom-left), normalized angular momentum (bottom-right). The bottom plots do not include the adapter section.}
    \label{fig:derbenevacc}
\end{figure}

\begin{figure}[tbp]
    \centering
    \includegraphics[width=0.49\linewidth]{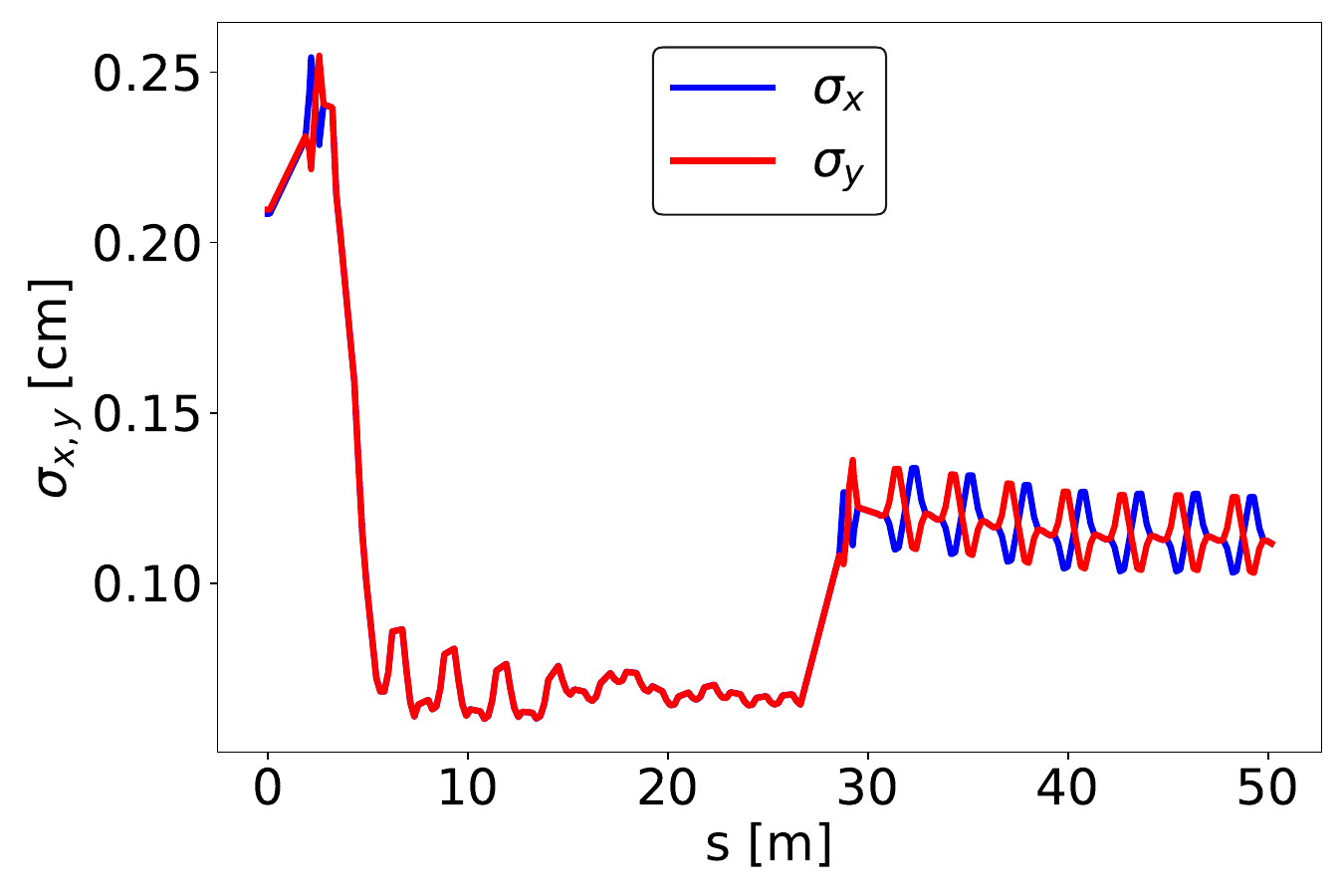}
    \includegraphics[width=0.49\linewidth]{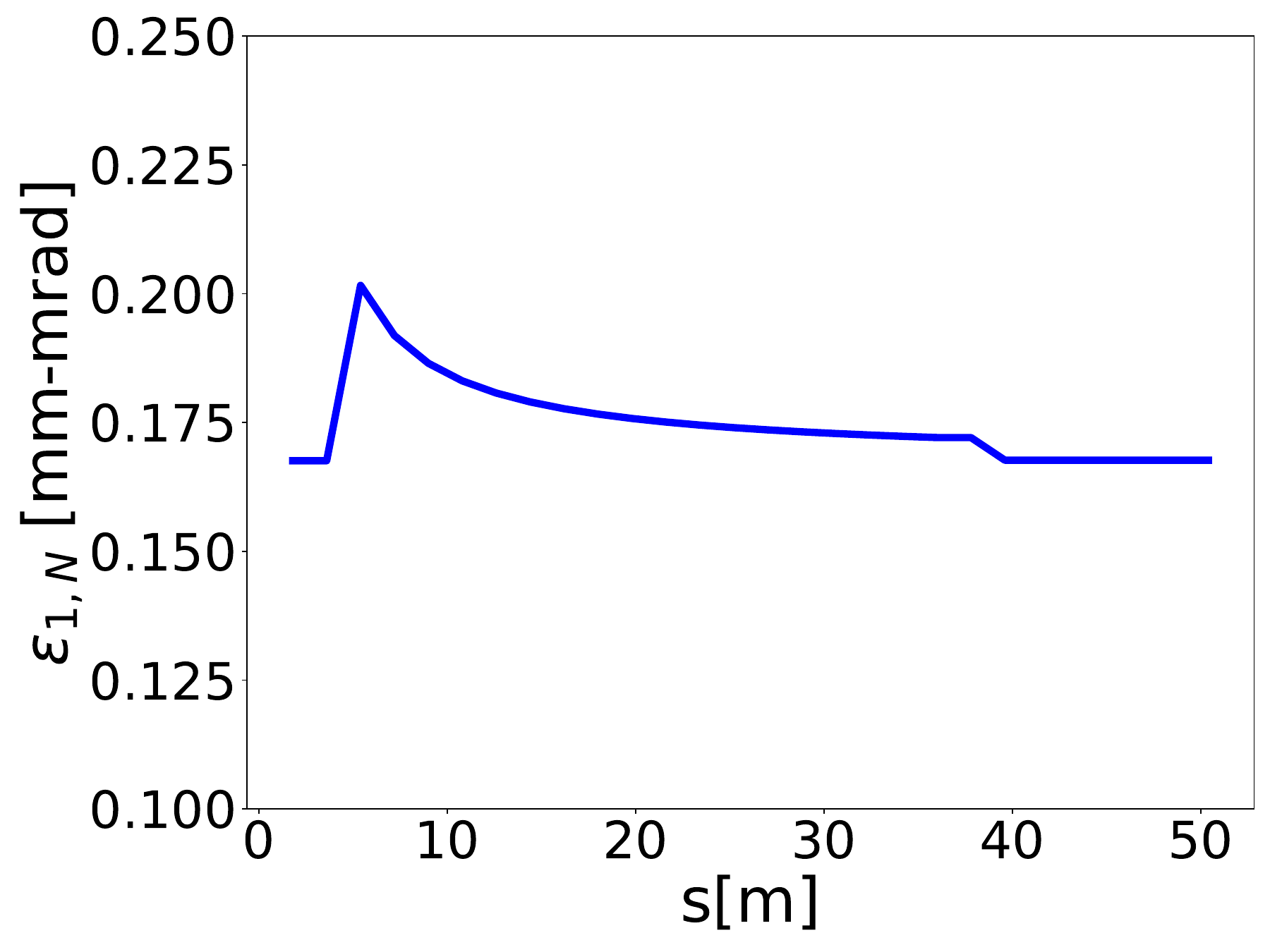}
    \includegraphics[width=0.49\linewidth]{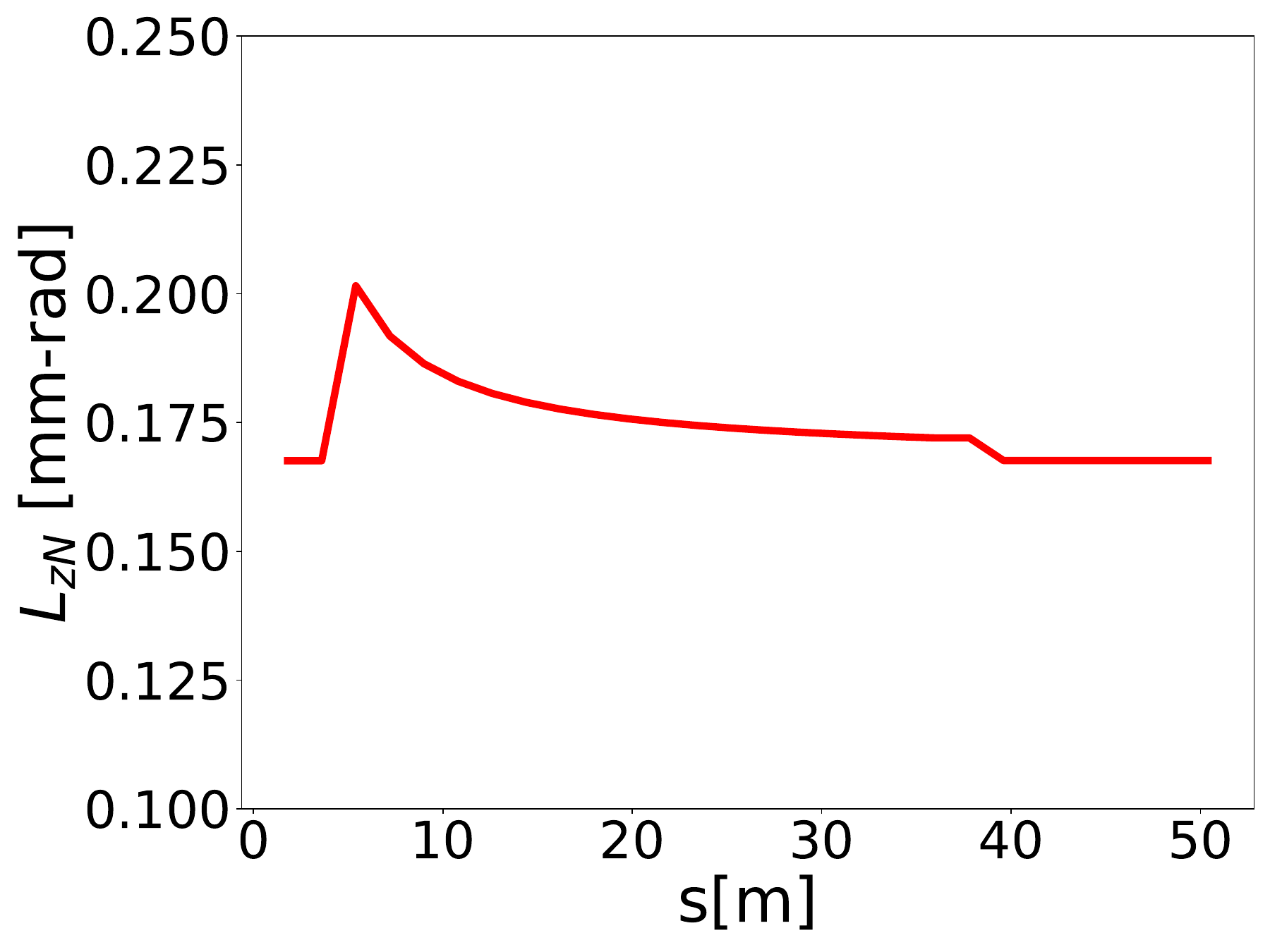}
    \includegraphics[width=0.49\linewidth]{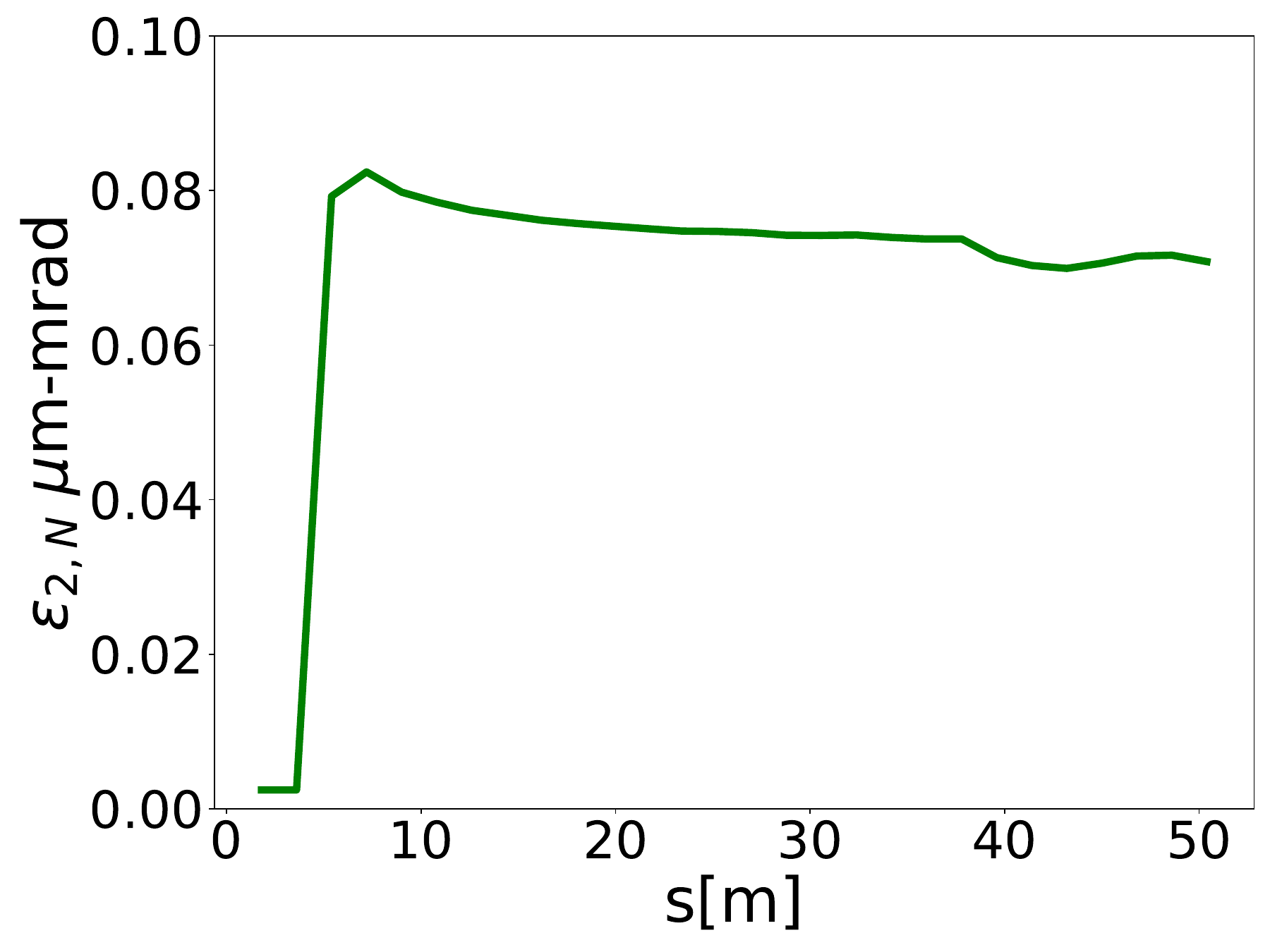}
    \caption{Beam parameters: RMS beam envelopes (top-left), RMS normalized eigenmode~1 emittance (top-right), RMS normalized angular momentum (bottom-left), RMS eigenmode~2 emittance (bottom-right).}
    \label{fig:halfsolacc}
\end{figure}

\begin{figure}[tbp]
    \centering
    \includegraphics[width=0.49\linewidth]{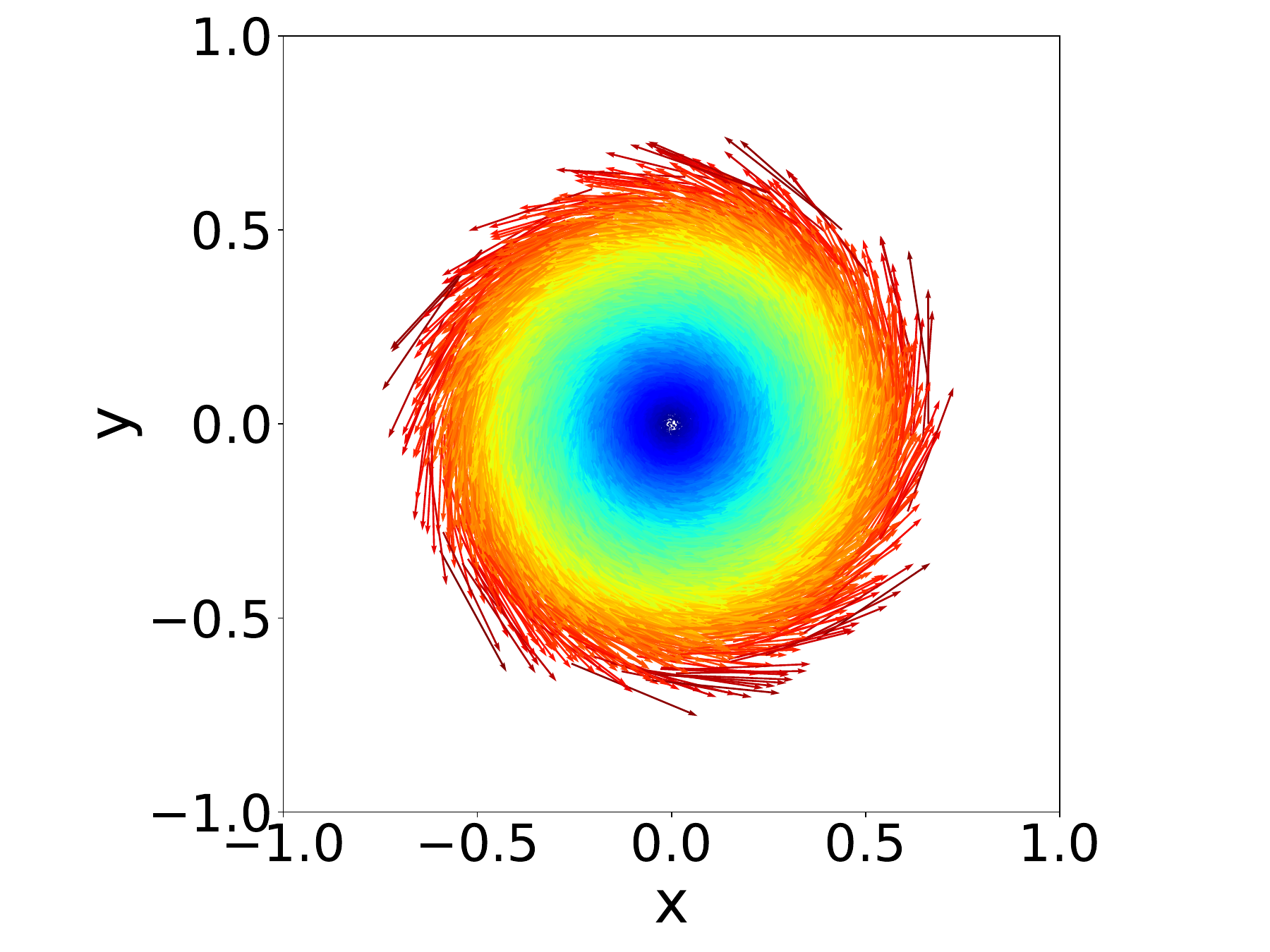}
    \includegraphics[width=0.49\linewidth]{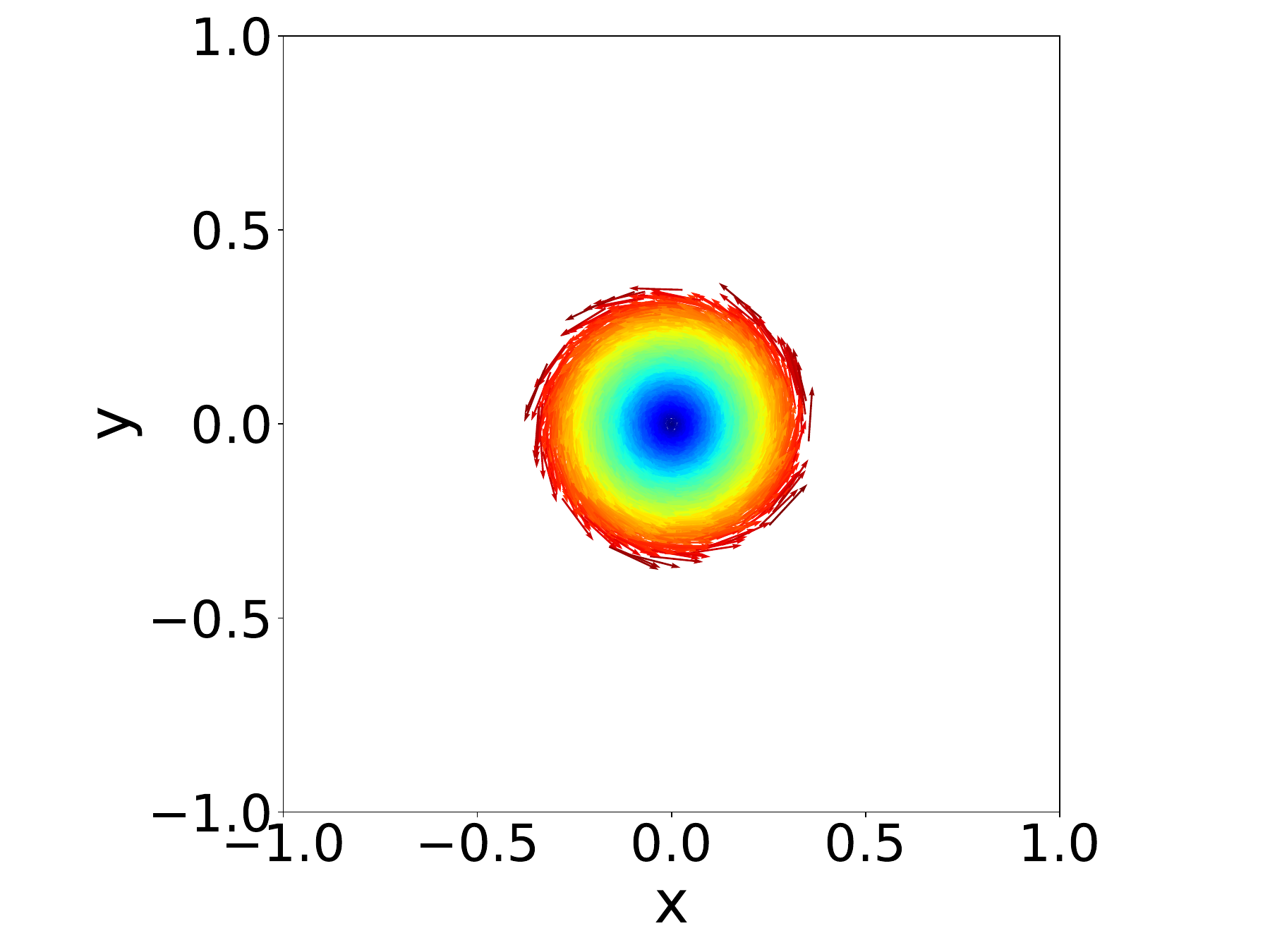}
    \caption{Circular Mode beam profile: Initial $(x,y)$ space (left), final $(x,y)$ space (right).}
    \label{fig:beamprof}
\end{figure}

In this section, we will show the acceleration process for a circular mode beam using the particle tracking code TRACK~\cite{TRACKweb} for simulating 2.5\,MeV protons. For the first case, we use Derbenev's adapter~\cite{derbenev1998adapting} to show the acceleration process with a solenoid channel. Parameters of this section are given in Table~\ref{tab:derbenev}.  
\begin{table}[btp]
    \centering
    \caption{First Case with Circular Beam from Derbenev's Adapter}
    \label{tab:derbenev}
    \begin{tabular}{cc}
    \toprule
         Parameters & Value  \\
    \midrule 
       $E_{k,i}$ [MeV] & 2.5 \\
       $E_{k,f}$ [MeV] & 20.0 \\
       $\epsilon_{0,N}$[mm$\cdot$mrad] & 0.073 \\
       $\beta_{x,y}$ [m] & 5.0 \\
       $\alpha_{x,y}$ & 0.0 \\
       $R=\epsilon_{1}/\epsilon_{2}$ & 686 \\ 
       \bottomrule
    \end{tabular}
\end{table}
Derbenev's adapter is used, which consists of three skew quadrupoles,  to transform an initial flat beam generated for protons to a circular mode. The process is illustrated by Fig.~\ref{fig:derbenevacc}. The cell containing one solenoid between two RF cavities is shown in the top-left plot of Fig.~\ref{fig:derbenevacc}.  
A perfect circular mode with proper magnetization requires the beam to have zero alpha functions. However, a magnetized beam does not necessarily produce a perfect circular mode beam. However, following the propagation of the phase of coupling from Eq.~\eqref{Eq:phaseofcouplingprop}, one can match it to circular mode beam optics. Based on this idea, Fig.~\ref{fig:halfsolacc} shows the production of the beam in the solenoid and later matched to the optics functions $\beta_{x,y}=5.0$\,m. It is then followed by an acceleration channel made of solenoids and RF cavities as shown in Fig.~\ref{fig:derbenevacc}. The beam is then matched back to different optics functions using four quadrupoles while maintaining the phase of coupling of $\pi/2$ after matching. The acceleration structure is then switched to quadrupole doublets and further accelerated to 52.5\,MeV. This shows that, considering the phase of coupling propagation, one can match the circular mode beam to different optics while preserving the angular momentum. The beam profile after magnetization and at the final section of the quadrupole channel is shown in Fig.~\ref{fig:beamprof}. Due to damping induced by acceleration and rematching of the optics the beam sizes are different. However, a clear indication of conserved angular momentum is seen, the arrows point in the direction of transverse momenta.

\section{Conclusion}
In conclusion, we have defined magnetized beams as circular mode beams through coupled optics. We have demonstrated that phase of coupling plays an important role in projection to various phase spaces. Defining circular modes through coupled optics, the propagation of phase of couplings became an important parameter in rotational invariant linac lattice designs. The condition of preserving the phase of coupling added extra constraints on design criteria which gives a clear method of how to match circular modes to a linac lattice. We have also shown that, through acceleration normalized angular momentum gets conserved similar to emittances. In the near future, we plan to run a full linac system that includes space charge effects as well which was left out from this paper.

\end{document}